\documentclass[12pt]{article}
\usepackage{makeidx,graphicx,epsf}
\newcommand{\ik}{{\it Kepler}}
\newcommand{\gtsimeq}{\raisebox{-0.6ex}{$\,\stackrel
        {\raisebox{-.2ex}{$\textstyle >$}}{\sim}\,$}}
\def\lesssim{\mathrel{\hbox{\rlap{\hbox{\lower4pt\hbox{$\sim$}}}\hbox{$<$}}}}
\def\gtrsim{\mathrel{\hbox{\rlap{\hbox{\lower4pt\hbox{$\sim$}}}\hbox{$>$}}}}
\def\ggg{\mathrel{\hbox{\rlap{\hbox{\lower4pt\hbox{$\sim$}}}\hbox{$>$}}}}

\begin{document}

{\bf{\large 
Kepler's Unparalleled Exploration of the Time Dimension
}}

This White Paper 
is submitted by the {\it Kepler Eclipsing Binary Star Working Group}
on 2013 September 3 in response to the 
\emph{``Call for White Papers: Soliciting Community Input for Alternate 
Science Investigations for the Kepler Spacecraft -- An open solicitation 
from the Kepler Project office at NASA Ames Research Center''}
made on August 2nd, 2013.\\
\hrule

{\small{
William Welsh (San Diego State Univ.),
Steven Bloemen (Radboud Univ.\ Nijmegen)$^{\dagger}$,
Kyle Conroy (Vanderbilt),
Laurance Doyle (SETI Institute),
Daniel C. Fabrycky (Univ.\ Chicago),
Eric B. Ford (Penn State Univ.)$^{\dagger}$,
Nader Haghighipour (Univ.\ Hawaii),
Daniel Huber (NASA Ames),
Stephen Kane (San Francisco State Univ.),
Brian Kirk (UKZN, Villanova),
Veselin Kostov (Johns Hopkins Univ.),
Kaitlin Kratter (Hubble Fellow, JILA and CU/NIST),
Tsevi Mazeh (Tel Aviv Univ.),
Jerome Orosz (San Diego State Univ.),
Joshua Pepper (Lehigh Univ.),
Andrej Pr{\v s}a (Villanova Univ.),
Avi Shporer (Caltech),
and
Gur Windmiller (San Diego State Univ.)
}} 


\hspace*{2.7in} {\bf{\large{Abstract}}}\\
We show that the {\it Kepler} spacecraft in two--reaction wheel mode 
of operation is very well suited for the study of eclipsing binary star 
systems. Continued observations of the {\it Kepler} field will provide 
the most enduring and long-term valuable science.
It will enable the discovery and characterization of eclipsing binaries 
with periods greater than 1 year -- these are the most important, 
yet least understood binaries for habitable-zone planet background
considerations. The continued mission will also enable the 
investigation of hierarchical multiple systems (discovered through 
eclipse timing variations), and provide drastically improved orbital 
parameters for circumbinary planetary systems.

{\bf{\large{1. Introduction and Motivation}}}\\
The spectacular success of the \ik\ is a result of the
Mission's four pillars:\\
1.\ Ultra high-precision photometry ($\sim$30 ppm for 12.5 mag in 6.5 
hours)\\
2.\ Simultaneous observations of very many stars ($\sim$170,000 stars)\\
3.\ Nearly continuous coverage ($\sim$90\% duty cycle on the same 
stars)\\
4.\ Very long duration ($\sim$4 years exploration of the 4th dimension)\\
The high precision is an obvious signature of \ik, but the other three 
aspects are equally important. Without them, the mission could not have 
been successful.

The original \ik\ Mission's goal is to determine the frequency and 
characteristics of exo-planets by surveying a large number of stars
and searching for planetary transits. 
Short period planetary and eclipsing binary (EB) systems are easy to 
detect since their transit and eclipse events are frequent. But for 
the more interesting longer-period systems, e.g., planets near the 
habitable zone, transits and eclipses are infrequent. These orbital 
periods are on the order of hundreds of days for an 
Earth+Sun-like system.
A few-year mission will not be able to detect a meaningful number 
of such events. A single transit/eclipse event is not very useful; a 
minimum of two are required to even estimate the period. Three events 
is considered a minimum for candidacy (unless part of a multi-object 
system), but 4 or more are needed to begin to untangle some of the 
complexities of the orbit, like eccentricity and precession. For 
planets or stars with orbital periods of a year or longer, this 
demands more than the current 4 years of data to carry out a full
investigation.

There are numerous long-period \ik\ Objects of Interest (KOIs) and EBs 
for which we have only a few eclipse events. \ik\ was able to discover 
these objects because of its unique many--star and ``long--look'' 
observing strategy. As we show below, {\it we can capitalize on \ik's 
fantastic scientific legacy by continuing the Mission}. 
Regrettably we cannot continue the hunt for Earth-size planets around
Sun-like stars, but we can continue the search for Earth-size planets
around small stars, for larger planets (in particular, those in the
habitable zone), and for EBs where even the degraded Kepler photometry 
can provide ample signal.
With only 2 functioning reaction wheels, \ik's guiding is not stable 
enough to allow ultra-precision capability. But, \ik\ has not lost the 
other 3 pillars of what made it great, provided it remains pointed at 
the same field.

If \ik's reaction wheels did function, there would be no question that 
the best place to point the telescope would be its original field.
And, if a new 
hypothetical \ik\ telescope were to be launched, it would take 4 
years just to catch up to where the original mission left off --- 
showing how exceptionally valuable the temporal baseline is. 
Larger and more sensitive missions can and will be launched. But those 
will not allow detecting long-period systems, for which there is no 
substitute for temporal information. Assuming the mission can continue 
for up to 2 more years, pointing to any other field(s) will gain no more 
than 2 years of data, of poorer quality than the already existing 4 
years of \ik\ data. Keeping \ik\ on the original field gives a total of 
6+ years of information --- reaching sensitivity in the temporal 
dimension that simply cannot be achieved with any current or planned 
missions. Six years of nearly continuous observations of the same stars 
would create a legacy that would last for generations.


{\bf{\large{2. Science Drivers and Goals}}}\\
{\bf{2.1 Eclipsing Binary Stars}}\\
Binary stars are a natural outcome of star formation, and indeed, for
stars $\geq 1M_\odot$, binaries are not the exception but the rule 
(Raghavan 2010, Kraus 2011).
Eclipsing binary stars are a very special subset of binaries and
are the cornerstone of stellar 
astrophysics: their unique geometry allows us to directly measure key 
stellar parameters -- radii, masses, temperatures, and luminosities. We 
can measure the masses and radii to a few percent (Andersen 1991; 
Torres et al.\ 2010), and with \ik\ data, down to 1\% or better 
(e.g.\ Bass et al.\ 2012). An 
ensemble of systems enables further modeling that then yields the 
statistical relations that are used to calibrate stars across the H-R 
diagram (Harmanec 1988), determine accurate distances (Guinan et al.\ 
1998), and study a range of intrinsic phenomena such as pulsations, 
spots, accretion disks, etc.\ (Olah 2007). Nearly every topic in 
astronomy benefits from a better calibration of stellar physics, and 
\ik\ is enabling a factor of 10x better determination of masses, 
radii, temperatures and luminosities. Moreover, our interpretation of 
exoplanet transits is intrinsically limited by our characterization of 
the host star.

{\bf{2.2 Binary Science Goals for an Extended Mission in the Kepler 
Field}}\\
We argue that continued monitoring of the \ik\ field will provide the 
highest impact science for a continued mission in two-reaction wheel 
mode -- which we refer to as {\it ``Kepler~II''}. In particular, it 
will allow the following unique achievements.

\pagebreak
{\bf{2.2.1 The discovery and characterization of EBs with $P>1$ year:}}
\ik\ has been incredibly fruitful for the study of binary stars; 
the \ik\ Eclipsing Binary Star Catalogs I, II, and III (Pr{\v s}a, et 
al.\ 2010, Slawson et al.\ 2011; Kirk et al.\ 2013) 
are major deliverables of the  \ik\ Mission. However, the investigation 
remains incomplete: the longer-period EBs are under-represented if not 
outright missing. Long-period binary systems are far more numerous in 
the sky: the field distribution is log-normal, peaking at $\sim50$ AU 
(Raghavan 2010). 
However, the eclipsing ones are observationally rare, due to the precise 
alignment needed between the observer and the binary orbital plane. In 
the current \ik\ EB Catalog there are 989 systems with 
periods between 0.001 and 0.01 years, 
848 systems between 0.01--0.1 years, 255 between 0.1--1.0 years, 
but only 14 between 1.0--10 years. There are many more long-period 
systems awaiting discovery in the \ik\ field if only we keep 
looking.

The discovery of long-period systems is invaluable for several reasons. 
First, long-period EBs are crucial for \ik's primary mission goal of 
determining $\eta$ Earth: these long-period eclipsing systems are the 
most important for estimating the occurrence rate of background EBs for 
determining the false-positive KOIs of habitable planets. Second, long 
period systems are particularly well suited for benchmarking stellar 
properties; one obtains all of the stellar parameters without the added 
complications of tidal interactions. Even though radial velocity (RV) 
surveys can partially characterize these systems, 
the precision of the stellar and orbital parameters will be far 
superior for systems that eclipse. And of course, eclipses provide 
radii, while RVs do not.

More broadly, a large sample of longer-period EBs can help resolve 
important unanswered questions in binary formation theory. Even with the 
torrent of new data, close binaries still present challenges to theories 
of binary formation (Artymowicz \& Lubow 1996, Bate 2000, Tohline 2002).  
There is no single mechanism that can explain the range of observed 
systems. The existence of planets in these systems further restricts 
formation pathways by setting a very stringent timescale on the host 
system's orbital evolution to small periods. While RV surveys can and do 
discover binaries in this period range, the light curves observable 
with \ik\ will allow us to measure the radii and stellar spin periods 
(via starspots), and also make best use of the Rossiter-McLaughlin (R-M) 
effect. The R-M effect provides the relative angle between the stellar 
angular momentum and the orbital angular momentum. Thus these data can 
uniquely distinguish between migration-based and dynamically-driven 
models for close binary formation.

{\bf{2.2.2 The discovery and characterization of hierarchical multiple 
systems:}}
Stellar and substellar tertiaries in binary systems are observed either 
directly (by detecting tertiary eclipse/transit events in the \ik\ 
light curve) or indirectly (from eclipse timing variations). By modeling 
eclipse shapes and dynamical aspects simultaneously -- via the method 
called photodynamical modeling -- the precision of derived fundamental 
parameters of the system can reach an astounding $\sim$0.2\% 
in radius and $\sim$0.5\% in mass
(Carter el al.\ 2011, Doyle et al.\ 2011), an order of magnitude better 
than what we can obtain from eclipsing binaries alone (Torres et al.\ 
2010). Thus multiple star systems are truly superior for stellar and 
orbital parameter calibration.

Detecting multiple systems is very challenging and thus it is no 
surprise that so many major discoveries are credited to \ik\ -- because 
of its long-term, uninterrupted observations of the same field. Temporal 
baseline is extremely important in this regard because tertiaries will 
always have comparatively long periods as required by dynamic 
stability of the system. In particular, 32 out of 111 short-period 
binaries that exhibit eclipse timing variations (ETVs) 
indicate a presence of a tertiary component with a period longer than 
4 years (Conroy et al.\ 2013)
meaning that a \emph{third} of the sample lacks sufficient temporal
coverage. 
For the longer-period EBs (P$\gtsimeq$1 day), the rate of triple-star
systems is 27\% (Orosz et al.\ 2013).
Continued surveillance of the \ik\ field, even at degraded 
photometric quality, is the only way to garner a statistically 
significant sample. Such a sample will also shed light onto formation 
theories by allowing for the study of changes in mass ratios and orbital 
properties with spectral type.

{\bf{2.2.3 Improved parameters (orbital and mass) for circumbinary 
planetary systems:}}
The degraded photometry of \ik\ will likely prohibit the discovery of 
new transiting circumbinary planets if their depths are comparable to 
those in the current sample (aside from Kepler-16 whose primary 
transits would be easy to detect). Nevertheless the continued 
monitoring of the existing 14 systems will provide far better 
constraints on the planetary parameters. For many of the detected 
circumbinary planets, there are more degrees of freedom in the 
dynamical modeling than there are transits. We expect to be able to 
detect some predicted future transits, and the transit timing 
information will enable much better determination of the planet's 
orbit. Deviations from the predicted time may indicate the presence
of non-eclipsing planets.
In addition, longer-duration monitoring will allow us to become 
sensitive to planets at larger semi-major axes from their host stars -- 
and these are predicted to be the giant planets (Pierens \& Nelson 
2008), and thus have ample transit depths for detection.
 

\begin{figure}[t]
{\includegraphics[scale=0.40,angle=-90]{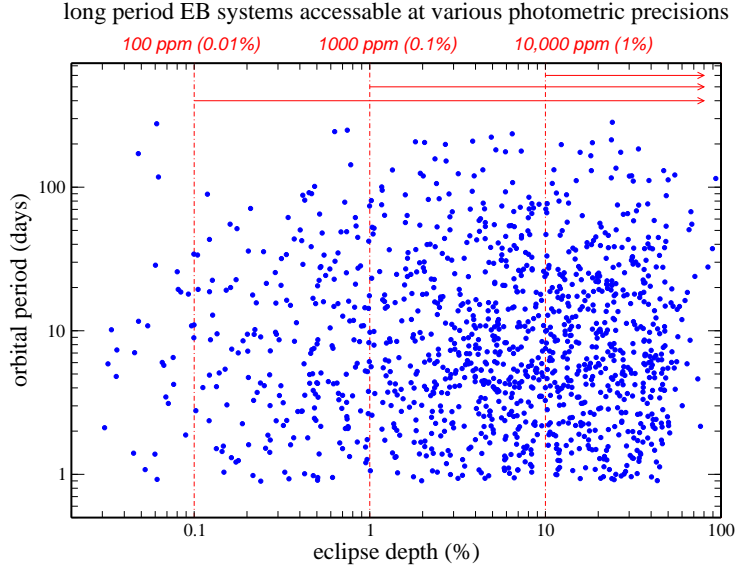}}
\hfill \parbox[t]{2.35in}{\caption[]{
{\textbf{EBs observable with {\it Kepler II} ---}}
The orbital period for the longer-period {\it Kepler} EB sample
is plotted versus the primary eclipse depth.
{\it Assuming a threshold of S/N of $>$10 for being useful,}
the dashed vertical lines show the photometric precision needed
to measure the eclipses for various eclipse depths.
}}
\end{figure}
{\bf{\large{3.\ Feasibility and Expected Photometric Performance}}}\\
{\bf{3.1~Feasibility of Proposed Goals:}}
To demonstrate the feasibility of our science goals, we consider the 
detectability of the current {\it Kepler} EB sample with a more noisy
{\it Kepler II} mission. Because we are mainly concerned with the 
longer-period  EBs in this White Paper, we make use of the Orosz, et 
al.\ (2013) sample of EBs with $P\gtsimeq 0.8$ days, but note that this 
is mildly incomplete due to on-going work: there are 24 systems with 
$P>1$ yr not yet analyzed in addition to the sample of 1250 EBs
shown in Figures 1--4.

Figure 1 shows the orbital period of the longer-period EB sample plotted 
against the primary eclipse depth. If we require the eclipse depth 
to be 10x larger than the short-term photometric noise, the dashed red 
vertical lines show the photometric precision needed to measure the 
eclipses times as a function of the eclipse depth. Points to the right 
of the dashed lines are measurable with the precision marked along the 
top of the figure. For example, if the eclipse depth is 1\%, a 
photometric precision of 0.1\% is required. Because the eclipses are so 
deep (median depth is 6.6\%), analysis of many, if not most, of the 
sample is possible even with significantly degraded photometric 
precision. Even if only 1\% precision is available, that leaves 509 EBs 
accessible to continued analysis in the {\it Kepler II} mission, or 
$\sim40\%$ of the long-period EB sample.

\begin{figure}[t]
{\includegraphics[scale=0.40,angle=-90]{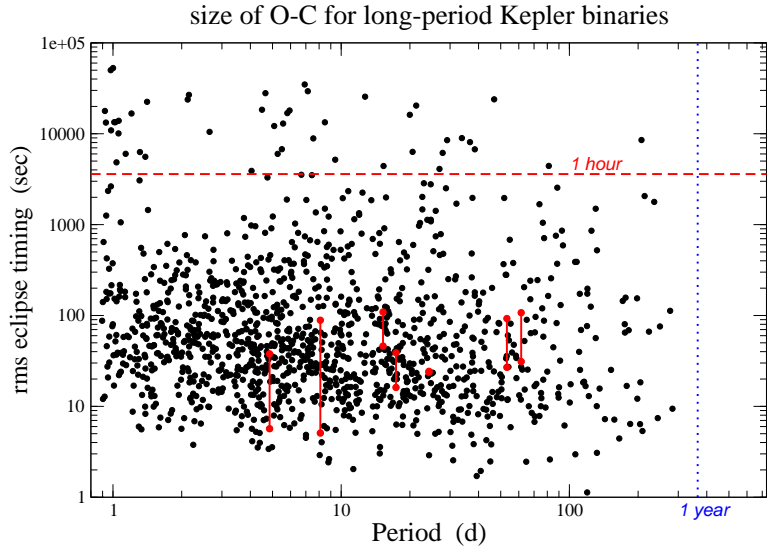}}
{\caption[]{
{\bf O-C vs P ---} 
The orbital period of a binary star will normally be constant, yielding
an O-C curve that is zero aside from noise. But if a third body 
perturbs the binary, the O-C will have systematic patterns and the
rms of the O-C will be large. The upper portion of this figure shows
those EBs with exceptionally large O-C variations, due to a
third (or more) star. The right-hand portion of the figure is empty,
showing the lack of long-period EBs.
The red points connected with a vertical line show the current rms
timing variations (lower points) and the expected rms for a
{\it Kepler II} mission (upper points).
}}
\end{figure}
Figure 2 shows the rms deviations of the primary eclipse times
from a linear ephemeris (i.e.\ the O-C amplitude) versus the binary 
period. The median period of the Orosz et al.\ (2013) sample 
is 7.13 days. Several important features are illustrated: (i) The median 
O-C rms is only 46.4 sec; by contrast, the points at the top of the 
figure have huge eclipse timing variations ($>1$ hour). These are not 
due to poor measurements; they are real variations that are caused by a 
third star perturbing the binary orbit. As noted above, these 
$\sim$ 50
systems are prime targets for an extended mission in the \ik\ field. 
(ii) These huge timing variations are so large that timing precision of 
even hundreds of seconds would still be more than adequate to help 
measure the properties of the third star. 
(iii) The right-hand part of the figure is sparse -- these are where 
the longest-period binaries would reside, and where we would gain the 
most from continuing in the original \ik\ field. Shorter surveys would 
simply re-populate the shorter-period part of the figure. (iv) The red 
points connected with a vertical line illustrate how the timing 
precision degrades (moves up) with the expected {\it Kepler II}
photometric performance (based on simulations described below). In 
some cases, the degradation is completely negligible. In other cases it 
is a factor of $\sim 20$ worse. For many cases, the timing is still 
excellent and sufficient for investigations of third body dynamics.

\begin{figure}[t] 
{\includegraphics[scale=0.40,angle=-90]{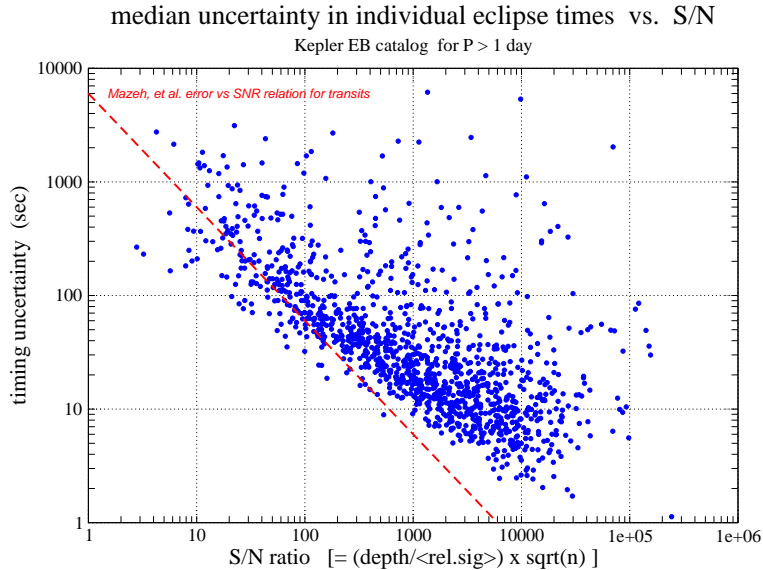} }
\hfill \parbox[t]{2.2in}{\caption[]{
{\bf Eclipse Timing Uncertainty ---}
The precision with which we measure eclipse times is shown as a
function of the signal-to-noise ratio. The SNR spans almost a factor 
of $10^5$, and the median precision is $<$ 30 sec.
A large number of EBs have such high SNR that a degradation of even
a factor of 50x in photometric precision will still allow timing 
precision to better than 100 sec.
}}
\end{figure} 
Figure 3 shows the median uncertainties in the measured eclipse times 
versus the signal-to-noise ratio (SNR). As before, these are for the 
longer-period EBs that have a detached or semi-detached morphology. 
Because the eclipse signal is so strong for these data (the median SNR 
is $\sim 1200$), the median uncertainty in a measurement of an 
individual eclipse time is only 28.9 sec, which is roughly a factor of 
20 better than the median uncertainty in planetary transit times. 
The expected trend, based purely on random-noise statistics, is
illustrated by the dashed line representing the transit-timing 
uncertainty relation from Mazeh et al.\ (2013): 
$\sigma_{TT} = $100/SNR (minutes). 
Scatter off this line is likely due to intrinsic stellar 
noise (starspots, pulsation, etc.). At very high SNR, the data deviate 
significantly from the expected line, suggesting the onset of a noise 
floor, perhaps caused by the 30-min cadence binning. This would imply 
that for these cases, as the SNR degrades due to poorer photometry, the 
loss in timing precision is not as steep as expected. This is born out 
in the simulations discussed  below.

\begin{figure}[h] 
{\includegraphics[scale=0.355,angle=-90]{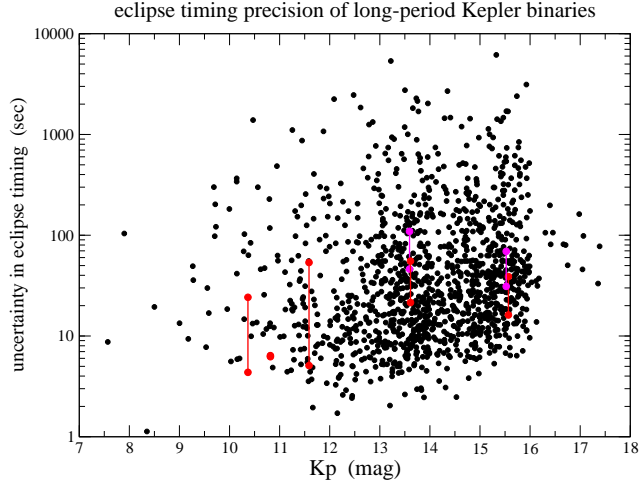} }
\hfill \parbox[t]{2.6in}
{\caption[]{
{\bf Eclipse Timing precision ---} The median uncertainty of 
individual eclipse timing measurements is shown versus the
Kepler magnitude of the star.
Seven test cases are shown in red, illustrating the expected 
change in timing precision due to the anticipated photometric 
degradation in the {\it Kepler II} mission.
}}

\end{figure} 
The same data can be plotted versus \ik\ magnitude, as shown in 
Figure~4. The sample has a median brightness of Kp = 13.96 mag. Note 
that the median timing precision is not a strong function of Kp: the 
precision is relatively flat out to 16th magnitude. This means that as 
the noise increases, the precision of the eclipse timing does not rise 
nearly as quickly as naively expected. While the timing uncertainty is 
not insensitive to the photometric noise, the expected degradation is 
not important for the higher-SNR cases or for the cases with large 
eclipse timing variations.

Finally, it is important to recall that much of the \ik\  mission's 
success has been due to the significant catalog preparation that 
preceded the mission, i.e., the KIC (Latham et al.\ 2005, Brown et al.\ 
2011, and later Pinsonneault et al. 2012), plus extensive follow-up 
observations (KFOP). 
By retaining the \ik\ field, we can build on: 
(1) all available auxiliary data already at hand, (2) all
\ik\ observations from the first 4 years of the mission that are of
unique photometric precision, and (3) the ongoing effort by the 
Community Follow-up Program (CFOP) to acquire follow-up observations.

{\bf{3.2~Simulated Expected Performance:}}
While the expected photometric performance of {\it Kepler II} when 
pointed at the original {\it Kepler} field is not known, a {\it rough} 
estimate can be made. Fortunately, even a rough estimate is sufficient 
to demonstrate that observations of eclipsing binaries will yield 
scientifically valuable information.

The dominant source of additional noise in the {\it Kepler II} 
photometry will be caused by pixel-to-pixel variations in sensitivity
in the CCD. Prior to the loss of the reaction wheels, the telescope 
guiding was very stable at sub-pixel levels. But without three reaction 
wheels, the guiding will drift by roughly 2 arcmin per day
(=0.625 pix per 30-minute cadence), and consequently the 1\%
imperfections in flat fielding will be manifested in the light curves.
We created a few simulated {\it Kepler II} light curves based on 
information made available by Ball Aerospace on 2013 Aug 20. 
Briefly, we degraded real {\it Kepler} light curves of eclipsing binaries 
using a noise model that includes the CCD flat field sensitivity 
variations. Due to the drift across pixels, the flat field noise is 
correlated across 2.5 hours, and this was simulated as a moving average 
(MA) process. A full description of the simulation is available 
at the Eclipsing Binary Catalog webpage:
http://keplerebs.villanova.edu/includes/appendix.pdf.
Using these simulated light curves, we measured the uncertainty on the
eclipse times.

As noted in Figure 3, the precision with which we can measure eclipse
times is not particularly well-determined from statistical 
considerations alone. Thus a handful of simulations were run to 
estimate the precision and degradation of our eclipse timing capability.
The precision with which we can measure the eclipse times are 
shown as the red points in Figures 2 and 4.
We selected seven systems that span a wide range of brightness, and
six of those were not in any way special: they have typical eclipse 
depths and typical intrinsic and instrumental variability. The seventh
case was a very high SNR circumbinary planet case. This example
shows no significant degradation because its eclipse is nearly 50\% 
deep. 

Figure 5 shows samples of the light curves for two of the 
seven simulations we ran. These are the two {\it worse} cases in terms 
of absolute timing precision (110 sec for KID 10659313), and 
degradation in timing precision (factor of 17.4x worse for KID 
10601579). The take-away message is that even with much worse
photometric performance, the eclipse signal is so strong that eclipse
timing can still be precisely measured for a large number of systems.
\begin{figure}[h!] 
{\includegraphics[scale=0.38,angle=-90]{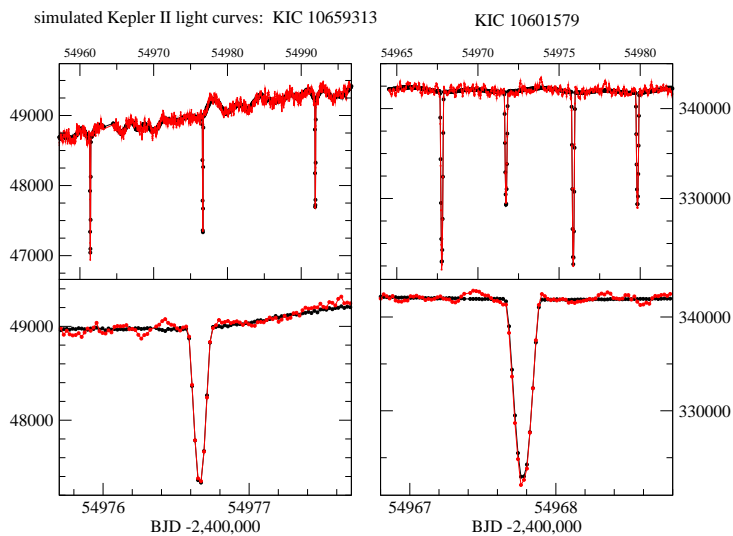} }
\hfill \parbox[t]{2.5in}
{\caption[]{
{\bf Example light curve degradation ---}
{\it{Upper panels:}} Simulated light curves of two typical eclipsing 
binaries. Original data is in black, degraded data in red. The 
correlated noise due to the spacecraft drift is apparent.
These two cases are the {\it worse} of our seven simulations. 
In the best cases, the noise is not visible on a scale that shows the 
full eclipse depth.
{\it{Lower panels}:} Close-up of upper panels. 
}}
\end{figure} 

{\bf{\large{4. Observing Mode Details}}}\\
{\bf 4.1 Focal plane mode:} Target apertures are needed. These must be 
large enough to capture the drifting starlight. 
{\bf 4.2 Cadence and Integration times:} If possible, a shorter cadence 
is strongly desired: we get a stronger signal (less smearing by 
convolution) and less noise (better pixel-to-pixel systematic noise 
removal). A shorter cadence means better resolution of sharp 
ingress/egress eclipse features, thus better analysis. To balance cadence
with sample size, 10 or 15 min cadence is desired.
{\bf 4.3 Data storage needs:} Since far fewer stars than the original 
mission are proposed, the data storage will generally not be a problem.
Even if {\it all} the KOIs and EBs were observed, this is only 
$\sim$6200 stars compared to the 170,000 currently observed. 
However, larger apertures are needed to accommodate the guiding drift. 
If the apertures are roughly 40 pixels long, then this very roughly 
takes 10x more memory. Then a 2x faster sampling
(i.e.\ 15 min cadence) would result in the same data storage needs as 
the original mission, and allow all KOIs, EBs, and $\sim$1300 other
targets to be observed. 
{\bf 4.4 Data Reduction:} While moving apertures are not needed, new
aperture positions are required for every spacecraft roll (i.e.\
daily). Since large apertures are needed (or contiguous sets of smaller 
ones), and the star is drifting within the aperture, the standard {\it 
Kepler} pipeline will not work. However, this is not nearly as 
challenging as it sounds: it is just like ground-based aperture 
photometry where you have to keep track of the star's x,y pixel position 
throughout the night and have a ``soft aperture'' within which to sum
the flux. The GO pixel-level photometry tools are the crux of the code.
What is needed is a way to track the optimal soft aperture as the star
drifts. Simple centroiding (just like in IRAF) is a good starting point.
{\bf 4.5 Target type:} Stellar point sources.
{\bf 4.6 Duration:} Targets should be observed continuously, for as long 
as possible.

{\bf{4.7 Highest Priority Eclipsing Binary Target List}}\\
$\bullet$ circumbinary planets: 14 systems \\
$\bullet$ long period EBs: 34 systems with P$>$ 300 days\\
$\bullet$ large ETVs: 280 systems (long-P and depth-changing EBs)\\
$\bullet$ large ETVs: 32 systems (short-P EBs)\\
$\bullet$ triply-eclipsing systems: 10  \\
$\bullet$ very low-mass EBs for precise M-R determination: 95 systems 
(Coughlin et al.~2011)\\
$\longrightarrow$ Bare minimum total number of EB systems: 465 

%

{\bf{4.8 Ground-based Eclipse Follow-up? No.}}\\
While observations from the ground are helpful for systems with short 
periods, there are very serious problems that makes such methods totally 
infeasible for the goals outlined in Section 2.2. Ground-based observing 
is interrupted by the diurnal cycle, seasons, and weather.  Those effects 
introduce the well-known observing window function (von Braun, et al. 
2009) which makes the discovery of long-period punctuated signals like 
transits and eclipses vanishingly small at periods much beyond one month.  
Furthermore, it is necessary to observe entire eclipses to characterize 
the systems described here.  The longer the period, the longer the 
eclipse duration, and once the duration exceeds one night, it becomes 
exceptionally hard to get full-eclipse coverage for more than a very few 
systems (multi-site campaigns are needed which often have significant 
systematics, and are both expensive in telescope time and risky due to 
weather). Finally, the most interesting cases are the ones where whole 
eclipses are impossible to predict within 12 hours due to the 
perturbations caused by the third body. Multi-site campaigns of several 
nights in duration would be needed for just one eclipse. While it might 
be possible to devote such resources to a few objects, it is not feasible 
for a statistically significant set of such eclipsing systems.



{\bf{\large{5. Arguments For Pointing Along the Ecliptic}}}\\
Strong arguments can be made for pointing {\it Kepler} at positions
along the ecliptic; but a stronger argument has been made to remain in 
the original field. Nevertheless, for completeness we list some 
advantages of moving to a field along the ecliptic.
1) The most significant advantage is the much better guiding stability 
and hence better photometric precision. However, for studies of 
eclipsing binaries, this is not that great an advantage, since the 
eclipse depths are so large that even several millimag precision is very 
useful.
2) Likely to be far less engineering work required, both for spacecraft 
management and for on-ground data calibration. This maps directly into 
significant savings in time and cost.
3) Given that roughly 1.5\% of all {\it Kepler} targets observed are 
EBs, we can expect hundreds to $\sim$2000 new EBs to be discovered. 
Some of these will be circumbinary planet hosts. The catalog of 
short-period EBs could conceivably be almost doubled. This would be 
impressive.
4) Several thousand new planet candidates will be identified; a great 
feat.
5) Discoveries of rare, exotic objects will be made.
6) Targeting a field that contains a well-studied open star cluster 
(e.g.\ the Hyades) would yield much better constraints on planet 
formation and evolution, since the planets would have the same age and 
composition.
7) With the $\sim$100-300 ppm precision expected if pointed along the 
ecliptic, asteroseismology of red-giant stars can be done, and more 
comparisons between asteroseismic- and EB--derived parameters can be 
made. Other variable stars will of course be found.

These are significant and exciting advantages, and it is abundantly clear 
that great science could be done if {\it Kepler} were pointed at fields
in the ecliptic. However, we must keep in mind that statistically, the 
objects {\it on average} will be the same (the exceptions being the 
youthful cluster stars), and the new study will not be as good as the 
original {\it Kepler} study (since the photometry is worse, and the
duration much shorter). We gain in numbers, and we gain on individual 
interesting objects, but we do not gain much in a Bayesian 
sense -- because of {\it Kepler} we have a strong prior on what to 
expect. It is where the prior is only weakly constrained, as in the 
longer temporal domain, that the information gain is maximized.
In addition, a {\it very} significant disadvantage of leaving the original 
{\it Kepler} FOV is the loss of the huge amount of information gathered 
on this field. It would take many years of effort to reproduce the
{\it Kepler Input Catalog} and all the Follow-Up Observations -- far
in excess of the effort needed to enable daily aperture 
rotations and the development of photometric measurement tools.
Unless abundant time and funding is available to reproduce the KIC and 
FOP, the loss of information is near catastrophic.  We conclude that 
while great science can be done along the ecliptic, even better science 
can be done in the original {\it Kepler} field.


{\bf{\large{6. References}}}$^{\dagger}$\\
{\small{
$\bullet$ {\it The Kepler Eclipsing Binary Catalog} -- third revision 
(beta): http://keplerebs.villanova.edu/ \\
$\bullet$ Appendix: 
http://keplerebs.villanova.edu/includes/appendix.pdf\\
Andersen, J. 1991, {\em A\&ARv}, {\bf 3}, 91\\
Artymowicz, P. \& Lubow, S.~H., 1996 {\em ApJL}, {\bf 467}, L77\\
Bass, G., et al.\ 2012, {\em ApJ}, {\bf 761}, 157\\
Bate, M.~R. 2000, {\em MNRAS}, {\bf 314}, 33\\
Carter, J. A., 2011, {\em Science}, {\bf 331}, 562\\
Conroy, K. E., et al.\ 2013, {\em AJ}, submitted\\
Coughlin et al.\ 2011, {\em AJ}, {\bf 141}, 78\\
Doyle, L. R., et al.\ 2011, {\em Science}, {\bf 333}, 1602\\
%
%
%
Guinan, E. F., et al.\ 1998, {\em ApJ}, {\bf 509}, L21\\
Harmanec, P. 1988, {\em BAICz}, {\bf 39}, 329\\
Holman, M. \& Weigert 1999, {\em AJ}, {\bf 117}, 621 \\ 
%
%
Kirk, B. et al.\ 2013, in preparation\\
Kraus, A.~L. et al.\ 2011 {\em ApJ}, {\bf 731}, 8 \\
Matijevi{\v c}, G., et al.\ 2012, {\em AJ}, {\bf 143}, 123\\
Ol\'ah, K. 2007, Proc. IAU Symp.\ \#240, eds.\ W.I. 
Hartkopf, E.F. Guinan, \& P. Harmanec, p. 442\\
Orosz, J.A., et al. 2013, in preparation$^{\dagger}$\\
Pierens, A. \& Nelson, R. P. 2008, {\em A\&A}, {\bf 483}, 633\\
Pinsonneault, M.H.\ et al. 2012, {\em ApJS} {\bf 199}, 30$^{\dagger}$\\
Pr{\v s}a, A., et al.\ 2011, {\em AJ}, {\bf 141}, 83\\
Raghavan, D., et al.\ 2010, {\em ApJS}, {\bf 190}, 1 \\
%
Sana, H.\ \& Evans, C.~J. 2010, {\em arXiv:1009.4197} \\
%
%
Slawson, R. W., et al.\ 2011, {\em AJ}, {\bf 142}, 160\\
Tohline, J.~E. 2002, {\em ARAA}, {\bf 40}, 349 \\
Torres, G., Andersen, J., \& Gim\'enez, A. 2010, {\em A\&ARv}, 
{\bf 18}, 67\\
Von Braun, K. et al.\ 2009, {\em ApJ}, {\bf 702}, 779$^{\dagger}$ 
}} 

$^{\dagger}$ indicates a minor addition/correction made to the white 
paper after submission.

\end{document}